\documentclass[aps,pra,twocolumn,superscriptaddress]{revtex4-2}
\usepackage{graphicx}
\usepackage{xcolor}
\usepackage{mathtools}
\usepackage{comment}
\usepackage{braket}
\usepackage{ulem}

\DeclareRobustCommand{\Erase}{\bgroup\markoverwith{\textcolor{red}{\rule[0.5ex]{2pt}{1.0pt}}}\ULon}
\begin{document}

\title{Trade-off relation between integrated metrological gain and local dissipation in magnetic-field sensing by quantum spin ensemble}

\author{Nozomu Takahashi}
\affiliation{Department of Applied Physics, Graduate School of Engineering, Tohoku University, Sendai 980-8579, Japan}

\author{Le Bin Ho} 
\thanks{binho@fris.tohoku.ac.jp}
\affiliation{Department of Applied Physics, Graduate School of Engineering, Tohoku University, Sendai 980-8579, Japan}
\affiliation{Frontier Research Institute 
for Interdisciplinary Sciences, 
Tohoku University, Sendai 980-8578, Japan}

\author{Hiroaki Matsueda}
\thanks{hiroaki.matsueda.c8@tohoku.ac.jp}
\affiliation{Department of Applied Physics, Graduate School of Engineering, Tohoku University, Sendai 980-8579, Japan}
\affiliation{
Center for Science and Innovation in Spintronics, Tohoku University, Sendai 980-8577, Japan
}
\date{\today}

\begin{abstract}
Quantum metrology plays a central role in precision sensing, where quantum enhancement of detection performance is crucial for both fundamental studies and practical applications. In this work, we derive a tight performance bound for magnetic-field sensing with a spin ensemble in the presence of dissipation. The metrological performance is quantified by the integrated metrological gain (IMG), which explicitly incorporates the time evolution of the measurement apparatus. By combining the Lindblad master equation with the quantum Fisher information, we obtain analytically exact trade-off relations between the IMG and the dissipation rate for local dephasing and local emission processes, showing that the gain scales inversely with the dissipation strength. This trade-off complements the Heisenberg limit, which addresses only the scaling with the number of spins and neglects dissipative dynamics. We analyze various initial state preparations and elucidate the role of quantum entanglement in the presence of dissipation. Notably, while entanglement is essential for achieving Heisenberg scaling at short times, it also accelerates dissipative degradation during time evolution. Consequently, for sufficiently long observation times, comparable metrological performance can be achieved even without entanglement.
\end{abstract}

\maketitle

\section{Introduction}

Quantum technology covers a wide range of research areas, including sensing, communication, and computation. Among these, quantum metrology is particularly promising for realistic sensing applications, even in the current noisy intermediate-scale quantum (NISQ) era \cite{Jiao}. Thus, a fundamental understanding of the mechanisms underlying quantum-enhanced metrological performance is essential. In this context, identifying the key factors that limit precision is crucial for clarifying the operational principles of quantum metrological protocols.

The minimum estimation error of a physical parameter is commonly bounded by the Heisenberg limit, which follows from information-theoretic arguments based on the Cram\'{e}r-Rao bound and the quantum Fisher information (QFI)~\cite{Braunstein,Paris,Giovannetti,Hyllus,Toth,Toth2,Pezze,Hervas}. This framework highlights the role of quantum fluctuations, relating the estimation precision to both the variance of the generator and the number of quantum resources employed. Related studies have examined precision limits in multiparameter estimation, highlighting the impact of noncommuting observables~\cite{Hou,Wang,Li}. Other work has shown that increasing precision does not always improve accuracy, even when unlimited resources are available~\cite{Song}.

Considering dissipative environments of NISQ devices and the accumulation of data from repeated measurements, it is essential to take into account the time evolution of measurement apparatus in the presence of dissipation, but research in this area is not sufficient~\cite{Kok,Pezze2,Escher,Demkowicz,Gammelmark,Liu,Li2}. In particular, an explicit trade-off between measurement precision and dissipation has yet to be fully established. Thus, we will address this issue in this work.

Research on measurement accuracy is also a hot topic in probabilistic thermodynamics and quantum thermodynamics. In particular, the thermodynamic uncertainty relation (TUR), first formulated in 2015, provides a thermodynamic expression of the widely recognized trade-off that higher precision requires higher energetic cost~\cite{Barato,Gingrich,Macieszczak}. The TUR establishes a lower bound on the ratio between the variance of an observable and the square of its mean, given by twice the inverse of the entropy production. It is possible to formulate the TUR as an estimation problem, and from that perspective, the TUR has a meaning very similar to the Cram\'{e}r-Rao inequality~\cite{Dechant,Dechant2,Dechant3,Hasegawa}. The speed limit theorem has also been reported to be a concept similar to TUR~\cite{Deffner,Vo,Hasegawa2}. 
More recently, TURs have been extended to continuous measurements and general open quantum systems~\cite{Hasegawa3,Brandner,Agarwalla,Hasegawa4}. The study of thermodynamic uncertainty in quantum systems still has potential for further development compared to classical cases. Thus, our study on the dynamical aspects of quantum metrology also provides new insights into quantum thermodynamics.

Motivated by both the technological development and fundamental scientific interests, we examine the trade-off relation between the performance of quantum metrology and the dissipation. A key quantity in this study is the integrated metrological gain (IMG), which provides a natural measure of metrological performance when the time evolution of the measurement apparatus is explicitly taken into account. Defined as the time integral over a measurement period, the IMG captures the competition between dissipative processes and the initial entanglement among spins in the measurement apparatus.

For this purpose, we combine the Lindblad equation with QFI and derive analytically exact trade-off relations between the IMG and the dissipation strength for both local dephasing and local emission processes, and examine these relations for various initial states. Our results clarify the role of quantum entanglement in the presence of dissipation and demonstrate that time-accumulated dissipative effects can counteract Heisenberg scaling, with initial entanglement in some cases becoming detrimental rather than beneficial.

The paper is organized as follows. In Sect.~\ref{LocalDephasing} and Sect.~\ref{LocalEmission}, we develop the theoretical framework for deriving the trade-off relation of the IMG in the presence of local dephasing and local emission noise, respectively. In Sect.~\ref{Discussion}, we discuss the physical implications of these results and summarize our conclusions.

\section{Trade-off relation for local dephasing noise}
\label{LocalDephasing}
\subsection{Model and General framework}

Let us start with our model. We are going to detect magnetic field by a spin ensemble. The Hamiltonian of the ensemble is defined by $H(\phi)=\phi J^z$, where $\phi$ is the strength of the magnetic field, $J^z$ is the total angular momentum $J^z=\sum_{j=1}^{N}\sigma_{j}^{z}/2$, and we consider $N$ independent spins without their interaction. Thus, the ensemble of these spins is embedded in the field, and we regard the ensemble as a field detector. We also consider the local dephasing that reduces the performance of quantum metrology.

In quantum metrology, QFI is a key parameter that obtains from the density matrix of the measurement apparatus. For this reason, let us start with the analysis of the density matrix. By starting with an initial density matrix $\rho_{\phi}(0)=\left|\psi_{\rm init}\right>\left<\psi_{\rm init}\right|$, the density operator at time $t$ satisfies the following Lindblad equation:
\begin{align}
\dot{\rho}_{\phi}(t)=&-i\left[H(\phi),\rho_{\phi}(t)\right] \nonumber \\
&+\sum_{\alpha}\gamma_{\alpha}\left[L_{\alpha}\rho_{\phi}(t)L_{\alpha}^{\dagger}-\frac{1}{2}\left\{L_{\alpha}^{\dagger}L_{\alpha}, \rho_{\phi}(t)\right\}\right], \label{Lindblad0}
\end{align}
where the dot denotes time derivative, $\alpha$ denotes each dissipation channel, $\gamma_\alpha$ is a dissipation rate, and $L_{\alpha}$ is the Lindbladian operator. For the moment, we consider local dephasing noise, and we omit $\alpha$-dependence of $\gamma_{\alpha}$. Then, the channel $\alpha$ is equal to each site $j$, and $L_{j}=\sigma_{j}^{z}$. The Lindblad equation is represented as
\begin{align}
\dot{\rho}_{\phi}(t)=-i\left[H(\phi),\rho_{\phi}(t)\right]-\gamma N\rho_{\phi}(t)+\gamma\sum_{j=1}^{N}\sigma_{j}^{z}\rho_{\phi}(t)\sigma_{j}^{z}.
\end{align}

According to Ref.~\cite{Nakazato}, we can solve this operator equation exactly. Let us introduce the operator $A=-iH(\phi)-\gamma N/2$, and then we obtain
\begin{align}
\dot{\rho}_{\phi}(t)=A\rho_{\phi}(t)+\rho_{\phi}(t)A^{\dagger}+\gamma\sum_{j}\sigma_{j}^{z}\rho_{\phi}(t)\sigma_{j}^{z}. \label{Lindblad}
\end{align}
By introducing the interaction picture for $A$, $\rho_{\phi}(t)=e^{At}\rho_{I}(t)e^{A^{\dagger}t}$ and by substituting it into Eq.~(\ref{Lindblad}), we obtain
\begin{align}
\dot{\rho}_{I}(t)&=\gamma\sum_{j}\left(e^{-At}\sigma_{j}^{z}e^{At}\right)\rho_{I}(t)\left(e^{A^{\dagger}t}\sigma_{j}^{z}e^{-A^{\dagger}t}\right) \nonumber \\
&=\gamma\sum_{j}\sigma_{j}^{z}\rho_{I}(t)\sigma_{j}^{z}.
\end{align}
Here, $A$ and $\sigma_{j}^{z}$ commute with each other, and the final result is greatly simplified. We may consider other types of dissipation in which the Lindbladian operators do not commute with the Hamiltonian, and in these cases the results would be much more complicated. However, our main conclusion about the trade-off relation does not change. Later, we will examine the local emission noise as one of other dissipation processes.

Let us introduce the basis state $\left|k\right>=\left|k_{1}k_{2}\cdots k_{N}\right>$ where $k_{j}$ ($j=1,2,...,N$) takes $0$ or $1$ depending on the spin state $\uparrow$ or $\downarrow$ at site $j$. By using this basis, the matrix element of $\rho_{I}(t)$ satisfies the following equation:
\begin{align}
\left<k\right|\dot{\rho}_{I}(t)\left|l\right>=\gamma\sum_{j}(-1)^{k_j+l_j}\left<k\right|\rho_{I}(t)\left|l\right>,
\end{align}
and the solution is given by
\begin{align}
\left<k\right|\rho_{I}(t)\left|l\right>=\left<k\right|\rho_{I}(0)\left|l\right>e^{\gamma t\sum_{j}(-1)^{k_j+l_j}}.
\end{align}
Then, the matrix element of $\rho_{\phi}(t)$ is given by
\begin{align}
\left<k\right|\rho_{\phi}(t)\left|l\right>=&\left<k\right|\rho_{\phi}(0)\left|l\right>e^{-\gamma t\left[N-\sum_{j}(-1)^{k_j+l_j}\right]} \nonumber \\
&\times e^{-(1/2)i\phi t\sum_{j}\left[(-1)^{k_j}-(-1)^{l_j}\right]}, \label{density}
\end{align}
where we used the relation $\rho_{I}(0)=\rho_{\phi}(0)$. Note that $tr\rho_{\phi}(t)=\sum_{k}\left<k\right|\rho_{\phi}(t)\left|k\right>=\sum_{k}\left<k\right|\rho_{\phi}(0)\left|k\right>=1$ for $\rho_{\phi}(0)=\left|\psi_{\rm init}\right>\left<\psi_{\rm init}\right|$.

The performance of the metrology is evaluated by the QFI defined as
\begin{align}
Q(t)=2\sum_{m,n}\frac{\left|\left<\psi_m(t)\right|\partial_{\phi}\rho_{\phi}(t)\left|\psi_n(t)\right>\right|^2}{\lambda_m(t)+\lambda_n(t)}, \label{QFI}
\end{align}
where $\psi_m(t)$ and $\lambda_m(t)$ are the eigenstate and the eigenvalue of $\rho_{\phi}(t)$, respectively. Equation (\ref{QFI}) is transformed into the following form with using Eq.~(\ref{density}):
\begin{align}
Q(t)=&2\sum_{m,n}\sum_{k,l}\frac{\left|\left<\psi_m(t)|k\right>\left<k\right|\partial_{\phi}\rho_{\phi}(t)\left|l\right>\left<l|\psi_n(t)\right>\right|^2}{\lambda_m(t)+\lambda_n(t)} \nonumber \\
=&\frac{1}{2}t^{2}\sum_{m\ne n}\frac{\left(\lambda_n(t)-\lambda_m(t)\right)^{2}}{\lambda_m(t)+\lambda_n(t)} \nonumber \\
&\;\;\;\;\;\;\times\left|\sum_{k}\left<\psi_m(t)|k\right>\sum_j(-1)^{k_j}\left<k|\psi_n(t)\right>\right|^2. \label{Qt2}
\end{align}
Note that the final result is also directly derived from the definitions of the interaction picture of $\rho_{\phi}(t)$ and $A$. In this case, we take the first derivative of $\rho_{\phi}(t)$ as $\partial_{\phi}\rho_{\phi}(t)=\partial_{\phi}\left(e^{At}\rho_{I}(t)e^{A^{\dagger}t}\right)=(it/2)\left[\rho_{\phi}(t),\sum_{j=1}^{N}\sigma_{j}^{z}\right]$, and substitute it to the definition of QFI.

According to our recent works~\cite{LBH,LBH2}, a key figure of merit for metrological performance is the metrological gain (MG), defined as $G(t)=Q(t)/t^{2}$, together with its time-integrated value evaluated over a sufficiently long interrogation period. Actually, we find the factor $t^{2}$ in Eq.~(\ref{Qt2}) and it is evidence for the factorization of $Q(t)$ by $t^{2}$. The integration of $G(t)$ over a sufficiently long period, $\mathcal{I}=\int_{0}^{\infty}G(t)dt$, is given by
\begin{align}
\mathcal{I}=&\frac{1}{2}\int_{0}^{\infty}dt\sum_{m\ne n}\frac{\big(\lambda_n(t)-\lambda_m(t)\big)^{2}}{\lambda_m(t)+\lambda_n(t)} \nonumber \\
&\;\;\;\;\;\;\times\left|\sum_{k}\left<\psi_m(t)|k\right>\sum_j(-1)^{k_j}\left<k|\psi_n(t)\right>\right|^2.
\end{align}
We call $\mathcal{I}$ an integrated metrological gain (IMG). An interesting observation is that the IMG has an inverse dimension of energy. Thus, this quantity serves as an effective estimator of energetic loss due to dissipation or of the energetic cost required to enhance metrological performance. Since the dissipation factor has a dimension of energy, this relation will give us useful information about their trade-off. In the following subsections, we examine the relationship between $\mathcal{I}$ and $\gamma$ for various initial conditions.

\subsection{Integrated metrological gain for GHZ-like initial state}

The Greenberger-Horne-Zeilinger (GHZ) state is often used as a good initial state of quantum metrology \cite{PhysRevLett.96.010401,Timm}. With this point in mind, we take $\left|\psi_{\rm init}\right>=\left(\left|00\cdots 0\right>+\left|11\cdots 1\right>\right)/\sqrt{2}$, which is an equal-weight superposition of the all-up and all-down states. Here, $\left|0\right>$ is spin-up state, and $\left|1\right>$ is spin-down. In this case, only four elements of $\rho_{\phi}(t)$ are non-zero: $\left<00\cdots 0\right|\rho_{\phi}(t)\left|00\cdots 0\right>=\left<11\cdots 1\right|\rho_{\phi}(t)\left|11\cdots 1\right>=1/2$ and $\left<00\cdots 0\right|\rho_{\phi}(t)\left|11\cdots 1\right>=\left<11\cdots 1\right|\rho_{\phi}(t)\left|00\cdots 0\right>^{\ast}=e^{-2N\gamma t}e^{-iN\phi t}/2$. Thus, it is possible to take a compact $2\times 2$ form of the density matrix spanned by the two bases $\left|00\cdots 0\right>$ and $\left|11\cdots 1\right>$. The non-zero eigenvalues and the corresponding eigenstates of $\rho_\phi(t)$ in the $2\times 2$ form are given by $\lambda_n(t)=\left(1+(-1)^{n}e^{-2N\gamma t}\right)/2$ and $\left<\psi_n\right|=\left(1,(-1)^{n}e^{iN\phi t}\right)/\sqrt{2}$, for $n=0,1$, respectively. Then, the MG is given by
\begin{align}
G_{\rm ent}(t)=N^{2}e^{-4N\gamma t}, \label{GN2}
\end{align}
where the subscript ``ent" stands for ``entangled" initial state. The appearance of the factor $N^2$ is clear evidence of the Heisenberg limit. The presence of the exponential damping factor with time originates in the dissipation rate $\gamma$. Furthermore, the exponential damping factor contains the factor $N$. Thus, even though the quantum enhancement due to the factor $N^{2}$ in front of the exponential term occurs, the factor $N$ also accelerates the exponential decay of MG in the time evolution. The IMG is obtained as
\begin{eqnarray}
\mathcal{I}_{\rm ent}=\int_{0}^{\infty}G_{\rm ent}(t)dt=\frac{N}{4\gamma}. \label{IMG}
\end{eqnarray}
Therefore, $\mathcal{I}$ decreases as the dissipation rate $\gamma$ increases. This is consistent with our recent numerical analysis \cite{LBH2}. The IMG is proportional to $N$, but does not depend on the strength of the magnetic field $\phi$. This result indicates that there is no competition between $\gamma$ and $\phi$, equivalently, dissipation affects all measurement fields in the same manner.

\subsection{Comparison of the integrated metrological gain between GHZ-like and product states}

For $N\ge 2$, the GHZ-like state introduced above is maximally entangled. To clarify the role of initial entanglement in metrological performance, we first consider the simplest nontrivial case $N=2$, and examine the product initial state, $\left|\psi_{\rm init}\right>=\left(\left|0\right>+\left|1\right>\right)/\sqrt{2}\otimes\left(\left|0\right>+\left|1\right>\right)/\sqrt{2}$. Then, the density matrix is given by
\begin{align}
\rho_{\phi}(t)=\frac{1}{4}\left(\begin{matrix}
1&e^{-2\gamma t}e^{-i\phi t}&e^{-2\gamma t}e^{-i\phi t}&e^{-4\gamma t}e^{-2i\phi t} \\
e^{-2\gamma t}e^{i\phi t}&1&e^{-4\gamma t}&e^{-2\gamma t}e^{-i\phi t} \\
e^{-2\gamma t}e^{i\phi t}&e^{-4\gamma t}&1&e^{-2\gamma t}e^{-i\phi t} \\
e^{-4\gamma t}e^{2i\phi t}&e^{-2\gamma t}e^{i\phi t}&e^{-2\gamma t}e^{i\phi t}&1
\end{matrix}\right).
\end{align}
We find that this density matrix is factorized into the product form $\rho_{\phi}(t)=\rho_{1}(t)\otimes\rho_{1}(t)$ with
\begin{align}
\rho_{1}(t)=\frac{1}{2}\left(\begin{matrix}1&e^{-2\gamma t}e^{-i\phi t}\\e^{-2\gamma t}e^{i\phi t}&1\end{matrix}\right). \label{factorized}
\end{align}
The most basic measure of mixed-state entanglement is the entanglement of formation~\cite{Bennett,Hill,Wootters}, $E\left(\rho_{\phi}(t)\right)={\rm min}\sum_{i}p_{i}E\left(\left|\psi_i\right>\right)$, where the minimization is taken over all pure-state decompositions $\rho_{\phi}(t)$, i.e., $\rho_{\phi}(t)=\sum_{i}p_{i}\left|\psi_i\right>\left<\psi_i\right|$ with all elements of states $\left|\psi_i\right>$ and probabilities $p_i$. Here, $E\left(\left|\psi_i\right>\right)$ is the bipartite entanglement entropy for the pure state $\left|\psi_i\right>$. Since $\rho_{\phi}(t)$ admits a decomposition into tensor products of the eigenstates of the reduced density matrix $\rho_1(t)$, the entanglement of formation vanishes at all times.

The eigenvalues of $\rho_{\phi}(t)$ are given by
\begin{align}
\lambda_{n}(t)&=\frac{1}{4}\left(1+(-1)^{n}e^{-2\gamma t}\right)^{2}, \\
\lambda_{2}(t)&=\lambda_{3}(t)=\frac{1}{4}\left(1-e^{-4\gamma t}\right),
\end{align}
where $n=0,1$ and the last eigenvalue is doubly degenerate. Their corresponding eigenvectors are given by
\begin{align}
&\left|\psi_n(t)\right>=\frac{1}{2}\left(\begin{matrix}1\\(-1)^{n}e^{i\phi t}\\(-1)^{n}e^{i\phi t}\\e^{2i\phi t}\end{matrix}\right) , \\
&\left|\psi_2(t)\right>=\frac{1}{\sqrt{2}}\left(\begin{matrix}1\\0\\0\\-e^{2i\phi t}\end{matrix}\right) , \left|\psi_3(t)\right>=\frac{1}{\sqrt{2}}\left(\begin{matrix}0\\1\\-1\\0\end{matrix}\right) .
\end{align}
The MG and IMG are given by
\begin{align}
G_{\rm sep}(t)&=2e^{-4\gamma t}, \label{Gproduct} \\
\mathcal{I}_{\rm sep}&=\frac{1}{2\gamma} \label{Iproduct},
\end{align}
respectively, where the subscript ``sep" stands for ``separable" initial state. 

Let us compare these results with the entanglement case in Eqs.~(\ref{GN2}) and (\ref{IMG}), where $G_{\rm ent}(t)=4e^{-8\gamma t}$ and $\mathcal{I}_{\rm ent}=1/2\gamma$ for $N=2$. The IMG does not depend on these two initial conditions. As we have already mentioned, Eq.~(\ref{GN2}) contains the factor $N^{2}$ but also the factor $N$ in the exponential decaying term. On the other hand, Eq.~(\ref{Gproduct}) does not contain a quantum enhancement factor originating the Heisenberg limit, but at the same time, the exponential decay due to the dissipation factor $\gamma$ is not strong.

The discussion above can be formulated more generally. Consider the case where the density matrix is factorized in the following form: $\rho_{\phi}(t)=\rho_{1}(t)\otimes\rho_{2}(t)\otimes\cdots\otimes\rho_{N}(t)$. The QFI is then transformed into
\begin{align}
Q(t)=&2\sum_{m_{1},...,m_{N}}\sum_{n_{1},...,n_{N}}\sum_{\alpha=1}^{N}\sum_{\alpha^{\prime}=1}^{N} \nonumber \\
&\frac{\left<\psi_{m_{\alpha}}\right|\partial_{\phi}\rho_{\alpha}\left|\psi_{n_{\alpha}}\right>\left<\psi_{n_{\alpha^{\prime}}}\right|\partial_{\phi}\rho_{\alpha^{\prime}}\left|\psi_{m_{\alpha^{\prime}}}\right>}{\prod_{i}\lambda_{m_{i}}+\prod_{j}\lambda_{n_{j}}} \nonumber \\
&\times\prod_{i\ne\alpha}\lambda_{m_{i}}\delta_{m_{i}n_{i}}\prod_{j\ne\alpha^{\prime}}\lambda_{m_{j}}\delta_{m_{j}n_{j}} \nonumber \\
=&2\sum_{\alpha}\sum_{m_{\alpha}}\sum_{n_{\alpha}}\frac{\left|\left<\psi_{m_{\alpha}}(t)\right|\partial_{\phi}\rho_{\alpha}(t)\left|\psi_{n_{\alpha}}(t)\right>\right|^{2}}{\lambda_{m_{\alpha}(t)}+\lambda_{n_{\alpha}}(t)},
\end{align}
where $m_{\alpha}$ and $n_{\alpha}$ ($\alpha=1,...,N$) take $0,1$, and the eigenstates and the eigenvalues are also factorized into their product forms: $\left|\psi_{m}(t)\right>=\left|\psi_{m_{1}}(t)\right>\otimes\left|\psi_{m_{2}}(t)\right>\otimes\cdots\otimes\left|\psi_{m_{N}}(t)\right>$, $\lambda_{m}(t)=\prod_{j=1}^{N}\lambda_{m_{j}}(t)$, and $\rho_{\alpha}(t)\left|\psi_{m_{\alpha}}(t)\right>=\lambda_{m_{\alpha}}(t)\left|\psi_{m_{\alpha}}(t)\right>$. Note that the off-diagonal terms for $\alpha$ and $\alpha^{\prime}$ vanish. Let us consider the case in which each spin obeys the independent Lindbladian dynamics with its own dissipation rate $\gamma_{\alpha}$. Then, the QFI is represented as
\begin{align}
Q(t)=t^{2}\sum_{\alpha=1}^{N}e^{-4\gamma_{\alpha}t}.
\end{align}
If all the dissipation rates are the same ($\gamma_{\alpha}=\gamma$), we obtain the following scaling
\begin{align}
Q(t)=t^{2}Ne^{-4\gamma t}.
\end{align}
The MG and IMG are then given by
\begin{align}
G_{\rm sep}(t)&=Ne^{-4\gamma t}, \label{Gsep} \\
\mathcal{I}_{\rm sep}&=\frac{N}{4\gamma}, \label{Isep}
\end{align}
respectively. As already mentioned, Eq.~(\ref{Isep}) is equal to Eq.~(\ref{IMG}). Let us also consider another case in which one of the dissipation rate is much smaller than the other ones, $\gamma_{0}\ll\gamma_{\alpha}$ ($\alpha\ne 0$). In this case, the QFI is approximately represented as
\begin{align}
Q(t)\sim t^{2}e^{-4\gamma_{0}t}.
\end{align}
The result suggests that when there are differences in spin relaxation, spins with faster relaxation do not contribute effectively to sensing.

\subsection{Ratio between Eqs.~(\ref{GN2}) and (\ref{Gsep})}

We compare Eq.~(\ref{GN2}) with Eq.~(\ref{Gsep}), and discuss whether the entangled initial state facilitates the metrological performance or not. For this purpose, we introduce their ratio as
\begin{align}
\frac{G_{\rm ent}(t)}{G_{\rm sep}(t)}=Ne^{-4(N-1)\gamma t}.
\end{align}
If this ratio is larger than one, the initial entanglement has some positive effects on the performance. This condition is represented as $N\ge e^{4(N-1)\gamma t}$. Taking logarithm of both the left and right hand sides leads to
\begin{align}
t\le\frac{1}{4}\frac{\log N}{N-1}\tau\le\frac{\tau}{4}.
\end{align}
Here, $\tau=\gamma^{-1}$ is a typical time scale of dissipation. The above inequality suggests that the time domain in which the initial entanglement dominates with respect to $G(t)$ is severely restricted as $N$ increases.

\subsection{Deviation from equal-weight superposition between $\left|0\right>$ and $\left|1\right>$ for $N=1$}

Here, we comment on how the MG changes for general initial states. For this purpose, we take $\left|\psi_{\rm init}\right>=\alpha\left|0\right>+\beta\left|1\right>$ ($\left|\alpha\right|^{2}+\left|\beta\right|^{2}=1$) for $N=1$. In this case, the density matrix is given by
\begin{align}
\rho_{\phi}(t)=\left(\begin{matrix}\left|\alpha\right|^{2}&\alpha\beta^{\ast}e^{-2\gamma t}e^{-i\phi t}\\ \alpha^{\ast}\beta e^{-2\gamma t}e^{i\phi t}&\left|\beta\right|^{2}\end{matrix}\right),
\end{align}
and the eigenvalues and the eigenstates are obtained as
\begin{align}
\lambda_{n}(t)&=\frac{1+(-1)^{n}\sqrt{\kappa}}{2}, \\
\left|\psi_{n}(t)\right>&=\dfrac{1}{\mathcal{N}_{n}}\left(\begin{matrix}1\\ \frac{e^{2\gamma t}e^{i\phi t}}{2\alpha\beta^{\ast}}\left(1-2\left|\alpha\right|^{2}+(-1)^{n}\sqrt{\kappa}\right)\end{matrix}\right),
\end{align}
where $\kappa=1-4\left|\alpha\right|^{2}\left|\beta\right|^{2}\left(1-e^{-4\gamma t}\right)$, $n=0,1$, and $\mathcal{N}_n$ is the normalization factor. The MG is given by
\begin{align}
G_{\rm 1}(t)=\frac{4\kappa}{\left|\mathcal{N}_{0}\mathcal{N}_{1}\right|^{2}},
\end{align}
Let us factorize $\alpha$ and $\beta$ as $\left|\alpha\right|^{2}=(1+\delta)/2$ and $\left|\beta\right|^{2}=(1-\delta)/2$, respectively. The parameter $\delta$ characterizes the deviation of $\left|\psi_{\rm init}\right>$ from the equal-weight superposition between $\left|0\right>$ and $\left|1\right>$. Then, the MG is transformed into the following form:
\begin{align}
G_{\rm 1}(t)=\left(1-\delta^{2}\right)e^{-4\gamma t}.
\end{align}
This is consistent with Eqs.~(\ref{GN2}) and (\ref{Gsep}), and the additional factor $1-\delta^2$ appears. We clearly find that the largest value of $G_{\rm 1}(t)$ is obtained for $\delta=0$. On the other hand, the MG becomes zero for $\delta=\pm 1$ ($\left|\psi_{\rm init}\right>$ is $\left|0\right>$ or $\left|1\right>$). This behavior is natural, since the magnetic field appears in the off-diagonal components of the density matrix. In this case also, the MG does not depend on the magnitude of the external magnetic field. 

\section{Trade-off relation for local emission noise}
\label{LocalEmission}
\subsection{Model and General framework}

In the previous section, we have considered local dephasing. Here, we discuss other dissipation processes, and show that our conclusion does not depend on the type of dissipation processes. As one of these processes, we consider local emission noise, and use $L_{j}=\sigma_{j}^{-}=\left|1\right>\left<0\right|$ in Eq.~(\ref{Lindblad0}). According to Ref.~\cite{Nakazato}, we again introduce the interaction picture $\rho_{\phi}(t)=e^{At}\rho_{I}(t)e^{A^{\dagger}t}$ for $A=-iH(\phi)-(1/2)\gamma\sum_{j}\sigma_{j}^{+}\sigma_{j}^{-}=-(1/2)i\phi\sum_{j}\sigma_{j}^{z}-(1/4)N\gamma-(1/4)\gamma\sum_{j}\sigma_{j}^{z}$. The differential equation for $\rho_{I}(t)$ is given by $\dot{\rho}_{I}(t)=\gamma\sum_{j}B_{j}(t)\rho_{I}(t)B_{j}^{\dagger}(t)$, where $B_{j}(t)=e^{-At}\sigma_{j}^{-}e^{At}=e^{-\gamma t/2}e^{-i\phi t}\sigma_{j}^{-}$. The equation for $\rho_{I}(t)$ is reduced to the following form:
\begin{align}
\dot{\rho}_{I}(t)=\gamma e^{-\gamma t}\sum_{j}\sigma_{j}^{-}\rho_{I}(t)\sigma_{j}^{+}. \label{rhoidot}
\end{align}

The solution of Eq.~(\ref{rhoidot}) is obtained by using the superoperator method. Let us define the superoperator as ${\cal L}_{j}\left[\rho\right]=\sigma_{j}^{-}\rho\sigma_{j}^{+}$. Then, the formal solution is given by
\begin{align}
\rho_{I}(t)&=\exp\left(\int_{0}^{t}d\tau\gamma e^{-\gamma\tau}\sum_{j=1}^{N}{\cal L}_{j}\right)\rho_{I}(0) \nonumber \\
&=\prod_{j=1}^{N}\exp\Bigg(\Lambda(t){\cal L}_{j}\Bigg)\rho_{I}(0) \nonumber \\
&=\prod_{j=1}^{N}\Bigg(I_{j}+\Lambda(t){\cal L}_{j}\Bigg)\rho_{\phi}(0), \label{superoperator}
\end{align}
where we used the identity ${\cal L}_{j}^{2}=0$, $\Lambda(t)=1-e^{-\gamma t}$, and $I_{j}$ is the identity operator. The density matrix is then obtained as $\rho_{\phi}(t)=e^{At}\rho_{I}(t)e^{A^{\dagger}t}$.

\subsection{Integrated metrological gain for GHZ-like initial state}

Let us consider the case of the GHZ-like initial state. Following Ref.~\cite{PhysRevA.107.042210}, we label the \(N\) spins by the set \(\Omega=\{1,2,\dots,N\}\). A configuration with \(n\) down spins is specified by a subset \(\alpha\subset\Omega\), where the cardinality \(|\alpha|=n\) denotes the number of elements in \(\alpha\). Using this notation, we define the computational basis state corresponding to $\alpha$ as
\begin{align}
    \ket{(N,n);\alpha}
\equiv
\bigotimes_{k\in\Omega\setminus\alpha}\ket{0_k}
\otimes
\bigotimes_{k\in\alpha}\ket{1_k},
\end{align}
that is, spins at sites $k\in \alpha$ are in the down state, while all remaining spins are in the up state. For example, \((N,n)=(3,2)\), we have 
\(
\Omega=\{1,2,3\}, \alpha\in\{\{1,2\},\{1,3\},\{2,3\}\}
\), and thus $\ket{(3,2);\{1,2\}}=\ket{110}$, $\ket{(3,2);\{1,3\}}=\ket{101}$, and $\ket{(3,2);\{2,3\}}=\ket{011}$.

Within this framework, the \(N\)-qubit GHZ state can be written compactly as a superposition of the two extremal excitation sectors,
\begin{align}
\ket{\mathrm{GHZ}_N}=
\frac{1}{\sqrt{2}}
\Big(
\ket{(N,0);\emptyset}
+
\ket{(N,N);\Omega}
\Big),
\end{align}
where \(\alpha = \emptyset\) denotes the empty set (no down spins) and \(\alpha = \Omega\) denotes the full set of sites (all spins down). Explicitly,
\begin{align}
    \ket{(N,0);\emptyset}=\ket{00\cdots0},
\ket{(N,N);\Omega}=\ket{11\cdots1}.
\end{align}
For these two elements, the subset labels \(\emptyset\) and \(\Omega\) are unique and therefore redundant. Accordingly, we will omit them and simply write \(\ket{(N,0)}\) and \(\ket{(N,N)}\) because no confusion can arise, with \(n=0\) and \(n=N\) serving as an implicit indicator of the corresponding configuration.

By using this basis, the GHZ-like initial state is represented as
\begin{align}
\rho_{\phi}(0)=&\frac{1}{2}\left(\left|(N,0)\right>\left<(N,0)\right|+\left|(N,0)\right>\left<(N,N)\right| \right. \nonumber \\
&\left. +\left|(N,N)\right>\left<(N,0)\right|+\left|(N,N)\right>\left<(N,N)\right|\right).
\end{align}
After the operation of the superoperators to $\rho_{\phi}(0)$, the non-zero components of $\rho_{I}(t)$ originate in the state $\left|(N,0)\right>\left<(N,0)\right|=\left|00\cdots 0\right>\left<00\cdots 0\right|$. For example, the product of $k$ superoperators, ${\cal L}_{j_1}{\cal L}_{j_2}\cdots{\cal L}_{j_k}$, produces the state $\left|(N,k);\alpha\right>\left<(N,k);\alpha\right|$, and the number of the states classified by the subset $\alpha$ is given by the combination of the $n$-down states, $c=\binom{N}{n}$. Then, we derive
\begin{align}
\rho_{I}(t)=&\rho_{\phi}(0)+\frac{1}{2}\sum_{n=1}^{N}\Lambda^{n}(t)\sum_{i=1}^{c}\left|(N,n);\alpha_i\right>\left<(N,n);\alpha_i\right|, \label{rhoitemission}
\end{align}
and the density matrix $\rho_{\phi}(t)$ is obtained as
\begin{align}
\rho_{\phi}(t)=&\frac{1}{2}e^{-N\gamma t}\left|(N,0)\right>\left<(N,0)\right| \nonumber \\
&+\frac{1}{2}e^{-N\gamma t/2}\left(e^{-iN\phi t}\left|(N,0)\right>\left<(N,N)\right|+{\rm h.c.}\right) \nonumber \\
&+\frac{1}{2}\left|(N,N)\right>\left<(N,N)\right| \nonumber \\
&+\frac{1}{2}\sum_{n=1}^{N}\Lambda^{n}(t)\sum_{i=1}^{c}e^{(n-N)\gamma t}\left|(N,n);\alpha_{i}\right>\left<(N,n);\alpha_{i}\right|,
\end{align}
where we used the following relation: $A\left|(N,n);\alpha_i\right>=\left[(-i\phi/2-\gamma/4)(N-2n)-N\gamma/4\right]\left|\left(N,n\right);\alpha_i\right>$. We can easily check that the trace of $\rho_{\phi}(t)$ is conserved.

The final result of $\rho_{\phi}(t)$ is block-diagonalized, and the $\phi$-dependent eigenvectors appear from the following $2\times 2$ block:
\begin{align}
\tilde{\rho}_{\phi}(t)
&=\frac{1}{2}\left(\begin{matrix}(1-\Lambda)^N & (1-\Lambda)^{N/2}e^{-iN\phi t}\\
(1-\Lambda)^{N/2}e^{iN\phi t} & 1+\Lambda^N\end{matrix}\right),
\end{align}
where the basis states of $\tilde{\rho}_{\phi}(t)$ are $\left|(N,0)\right>$ and $\left|(N,N)\right>$. As shown below, the $\phi$-independent eigenvectors do not contribute to QFI. The eigenvalues and the corresponding eigenstates of $\tilde{\rho}_{\phi}(t)$ are given by
\begin{align}
    \lambda_\pm(t)&=\frac{1}{4}\left(u_+ \pm\ \Delta\right),\\
    \Delta&=\sqrt{u_-^2+4(1-\Lambda)^N}, \\
    u_\pm &= 1+\Lambda^N\pm (1-\Lambda)^N,
\end{align}
and 
\begin{align}
\ket{\psi_\pm(t)}=
\frac{1}{\mathcal{N}_\pm(t)}
\left(\begin{matrix}
2(1-\Lambda)^{N/2}\\
e^{iN\phi t}\left(u_{-}\pm\Delta\right)
\end{matrix}\right),
\end{align}
where the normalization factors are $
\mathcal{N}_\pm^2(t)=2\Delta(\Delta\pm u_-).
$

The QFI is represented by
\begin{align}
Q(t)&=2\sum_{m\ne n}\frac{\left(\lambda_{m}(t)-\lambda_{n}(t)\right)^{2}}{\lambda_{m}(t)+\lambda_{n}(t)}\left|\left<\psi_{m}(t)|\partial_{\phi}\psi_{n}(t)\right>\right|^{2} \\
&=4\frac{\left(\lambda_{+}(t)-\lambda_{-}(t)\right)^{2}}{\lambda_{+}(t)+\lambda_{-}(t)}\left|\left<\psi_{+}(t)|\partial_{\phi}\psi_{-}(t)\right>\right|^{2},
\end{align}
and the MG is given by
\begin{align}\label{eq:Gent_emi}
G_{\rm ent}(t)&=2N^{2}\frac{(1-\Lambda)^{N}}{1+\Lambda^{N}+(1-\Lambda)^{N}} \nonumber \\
&=2N^{2}\frac{e^{-N\gamma t}}{1+\left(1-e^{-\gamma t}\right)^{N}+e^{-N\gamma t}}.
\end{align}
In this case, $G_{\rm ent}(t)$ does not show a single exponential form, but the asymptotic form of $G_{\rm ent}(t)$ is $N^{2}e^{-N\gamma t}$. Thus, the increase in $N$ also prohibits the metrological performance, and we have confirmed that our conclusion does not change even for this type of dissipation.

Let us evaluate the IMG by integrating $G_{\rm ent}(t)$ over time. Set \(x=e^{-\gamma t}\),
i.e., \(x\in(0,1]\) as \(t:0\to\infty\), and \(dt=-\frac{dx}{\gamma x}\), we have $e^{-N\gamma t}=x^N,\ 1-e^{-\gamma t}=1-x$. Then, the IMG is given by
\begin{align}\label{eq:I_ent_em_full}
\mathcal{I}_{\rm ent}\equiv\int_0^\infty G_{\rm ent}(t)dt = \frac{2N^2}{\gamma}\int_0^1
\frac{x^{N-1}}{1+(1-x)^N+x^N}dx,
\end{align}
and we find that the IMG scales inversely with $\gamma$. First, we note that \((1-x)^N\ge 0, \forall x\in[0,1)\), and  the denominator gives
\(
1+\left(1-x\right)^N+x^N\ge 1+x^N.
\)
Therefore
\begin{align}
\mathcal{I}_{\rm ent} \le \frac{2N^2}{\gamma}\int_0^1
\frac{x^{N-1}}{1+x^N}dx.
\end{align}
Then, the IMG is
\begin{align}\label{eq:I_ent_em_up}
\mathcal{I}_{\rm ent} \le
\frac{2N\ln 2}{\gamma},
\end{align}
where we set $u = x^N$ then used \(\int_0^1 \frac{du}{u+1}=\ln 2\).

For large $N$, 
we have 
\((1-x)^N\approx 0.
\)
Substituting this approximation into Eq.~\eqref{eq:I_ent_em_full}, we obtain
\begin{align}
\mathcal{I}_{\rm ent}
\approx
\frac{2N\ln 2}{\gamma},
\end{align}
which reaches the upper bound given in Eq.~\eqref{eq:I_ent_em_up}. As a consistency check, we may also estimate the IMG using the asymptotic decay of the gain
\(G_{\rm ent}(t)\approx N^2 e^{-N\gamma t}\), which captures the dominant long-time contribution. This yields
\begin{align}
{\cal I}_{\rm ent}\approx\frac{N}{\gamma}.
\end{align}

Note that the factor $4$ that appears in Eqs.~(\ref{GN2}) and (\ref{IMG}) for the local dephasing noise is missing in the above asymptotic results. We emphasize that, for the same dissipation rate \(\gamma\), local emission yields a larger information gain than local dephasing. In the emission case, the system evolves from the coherent superposition \((|0\rangle+|1\rangle)/\sqrt{2}\) toward the pure state \(|1\rangle\), thereby retaining population information. In contrast, local dephasing drives the system toward the maximally mixed state, leading to a complete loss of phase coherence and, consequently, of metrological information.

\subsection{Integrated metrological gain for product initial state and its comparison with the result for entangled initial state}

Let us also consider the case of the product initial state: $\rho_{\phi}(0)=\rho_{1}\otimes\rho_{2}\otimes\cdots\otimes\rho_{N}$, where $\rho_{j}=\left(\left|0\right>\left<0\right|+\left|0\right>\left<1\right|+\left|1\right>\left<0\right|+\left|1\right>\left<1\right|\right)/2$. Since each superoperator at site $j$ acts independently on each local density matrix $\rho_{j}$, we find
\begin{align}
\rho_{I}(t)=\bigotimes_{j=1}^{N}\left(I_{j}+\Lambda{\cal L}_{j}\right)\rho_{j}=\bigotimes_{j=1}^{N}\frac{1}{2}\left(\begin{matrix}1&1\\1&1+\Lambda\end{matrix}\right).
\end{align}
Then, the density matrix $\rho_{\phi}(t)$ is obtained as
\begin{align}
\rho_{\phi}(t)=\bigotimes_{j=1}^{N}\frac{1}{2}\left(\begin{matrix}e^{-\gamma t}&e^{-\gamma t/2}e^{-i\phi t}\\e^{-\gamma t/2}e^{i\phi t}&2-e^{-\gamma t}\end{matrix}\right).
\end{align}
In this case, the QFI, MG, and IMG are given respectively by
\begin{align}
Q(t)&=t^{2}Ne^{-\gamma t}, \\
G_{\rm sep}(t)&=Ne^{-\gamma t}, \\
{\cal I}_{\rm sep}&=\frac{N}{\gamma}.
\end{align}
According to the asymptotic form of $G_{\rm ent}(t)$, the IMG is nearly independent of the initial condition in the case of the local emission noise. We also find $G_{\rm ent}/G_{\rm sep}\simeq Ne^{-(N-1)\gamma t}$, and the advantage due to entanglement appears at the time range of $\gamma t<\log N/(N-1)$, which is severely limited.

\section{Discussion and Summary}
\label{Discussion}
Finally, we discuss implications of the present results and future prospects. Our results clearly show the trade-off relation between $\mathcal{I}$ and $\gamma$, but $\mathcal{I}$ does not contain information about $\phi$. It is naively considered that if the disorder dominates over the physical quantity being measured, the measurement becomes extremely difficult. In this respect, the magnitude of $\phi$ and the dissipation rate $\gamma$ may show some competitive relationship. We cannot find such a relationship in the present analysis. 

For comprehensive analysis, it is necessary to consider the measurement protocol, i.e., a specific positive operator-valued measure (POVM). In relation to this discussion, by combining the Cram\'{e}r-Rao inequality with the definition of $\mathcal{I}$, we can estimate the lower bound of $\mathcal{I}$ as follows:
\begin{align}
\mathcal{I}\ge\int_{0}^{\infty}\frac{dt}{t^{2}V_{\phi}(t)},
\end{align}
where $V_{\phi}(t)$ is the variance of the magnetic field strength $\phi$ and it should also depend on $\gamma$ through time-evolution processes. It is an interesting question to solve how the right hand side of this inequality behaves.

The present results can also be applicable to the case of energy charge to the measurement apparatus for keeping or enhancing its performance. The continuous energy charge can be represented by simply reversing the sign of the dissipation rate $\gamma$ in the Lindblad equation, although it is necessary to take appropriate charging processes as alternative Lindblad operators. Then, the metrological gain increases exponentially with time according to Eq.~(\ref{GN2}). In this case, we can easily expect that if the energy-charging rate exceeds with the dissipation rate the gain increases exponentially.

Recently, there are experimental reports on nanoscale sensing with nitrogen vacancy (NV) centers in diamond, and in these experiments the entangled NV pairs are used~\cite{Zhou,Rovny}. In the present study, we focused solely on the presence or absence of initial entanglement, so investigating the effects of entanglement between spins remains an interesting task for future research.

In summary, we have developed the theory for the trade-off relation between the performance of quantum metrology and the dissipation in terms of the integrated metrological gain. We have considered magnetic-field sensing with a spin ensemble, and have also taken into account the dissipation. We have analytically derived the exact formula for the trade-off relation between IMG and the dissipation rate in cases of local dephasing and emission noises. We have found that the IMG is inversely proportional to the dissipation rate. At least in the case of local dissipation noise, the IMG does not depend on whether the initial state is entangled or not. This is because the entanglement more quickly reduces MG through time evolution processes. Our result serves a valuable figure of merit of metrological performance, and thus provides quite useful information to the field of quantum metrology.

\acknowledgements
We acknowledge Ryo Tatsumi and Hiroki Morishita for discussion. We acknowledge financial support from KAKENHI projects No. JP24K00563, No. JP24K02948, No. JP24K06878, and No. JP23K13025. HM also acknowledges support from CSIS, Tohoku University. LBH is supported by the Tohoku Initiative for Fostering Global Researchers for Interdisciplinary Sciences (TI-FRIS) of MEXT's Strategic Professional Development Program for Young Researchers.

\section*{Data availability}
No data were created or analyzed in
this study.

\bibliography{refs}

\end{document}